\documentclass[11pt]{ceb-l}

\usepackage{amssymb}

\usepackage{graphicx}

\def\C{\mathcal{C}}

\def\od{\stackrel{\mathrm{def}}{=}}

\def\O{\mathcal{O}}

\def\RR{\mathbb{R}}
\def\U{\mathcal{U}}
\def\N{\mathcal{N}}

\newtheorem{theorem}{Theorem}[section]
\newtheorem{lemma}[theorem]{Lemma}

\theoremstyle{definition}
\newtheorem{definition}[theorem]{Definition}
\newtheorem{example}[theorem]{Example}

\theoremstyle{remark}

\numberwithin{equation}{section}

\begin{document}

\title[Topology of neural codes]{What can topology tell us about the\\ neural code?}


\author{Carina Curto}
\address{Department of Mathematics, The Pennsylvania State University}
\curraddr{}
\email{ccurto@psu.edu}
\thanks{}


\date{May 1, 2016}

\begin{abstract}
Neuroscience is undergoing a period of rapid experimental progress and expansion.  New mathematical tools, previously unknown in the neuroscience community, are now being used to tackle fundamental questions and analyze emerging data sets.  Consistent with this trend, the last decade has seen an uptick in the use of topological ideas and methods in neuroscience.  In this talk I will survey recent applications of topology in neuroscience, and explain why topology is an especially natural tool for understanding neural codes.  {\it Note: This is a write-up of my talk for the Current Events Bulletin, held at the 2016 Joint Math Meetings in Seattle, WA.}
\end{abstract}

\maketitle


\section{Introduction}
Applications of topology to scientific domains outside of pure mathematics are becoming increasingly common.  Neuroscience, a field undergoing a golden age of progress in its own right, is no exception.   The first reason for this is perhaps obvious -- at least to anyone familiar with topological data analysis.  Like other areas of biology, neuroscience is generating a lot of new data, and some of these data can be better understood with the help of topological methods.  A second reason is that a significant portion of neuroscience research involves studying networks, and networks are particularly amenable to topological tools.  Although my talk will touch on a variety of such applications, most of my attention will be devoted to a third reason -- namely, that many interesting problems in neuroscience contain topological questions in disguise.  This is especially true when it comes to understanding {\it neural codes}, and questions such as: how
do the collective activities of neurons represent information about the outside world?

I will begin this talk with some well-known examples of neural codes, and then use them to illustrate how topological ideas naturally arise in this context.  Next, I'll take a brief detour to describe other uses of topology in neuroscience.  Finally, I will return to neural codes and explain why topological methods are helpful for studying their intrinsic properties.  Taken together, these developments suggest that topology is not only useful for analyzing neuroscience data, but may also play a fundamental role in the theory of how the brain works.

\section{Neurons: nodes in a network or autonomous sensors?}

It has been known for more than a century, since the time of Golgi and Ramon y Cajal, that the neurons in our brains are connected to each other in vast, intricate networks.  Neurons are electrically active cells.  They communicate with each other by firing action potentials (spikes) -- tiny messages that are only received by neighboring (synaptically-connected) neurons in the network.  Suppose we were eavesdropping on a single neuron, carefully recording its electrical activity at each point in time.  What governs the neuron's behavior?  The obvious answer: it's the network, of course!
If we could monitor the activity of all the other neurons, and we knew exactly the pattern of connections between them, and were blessed with an excellent model describing all relevant dynamics, then (maybe?) we would be able to predict when our neuron will fire.  If this seems hopeless now, imagine how unpredictable the activity of a single neuron in a large cortical network must have seemed in the 1950s, when Hodgkin and Huxley had just finished working out the complex nonlinear dynamics of action potentials for a simple, isolated cell \cite{rinzel1990HodgkinHuxley}.

And yet, around 1959, a miracle happened.  It started when Hubel and Wiesel inserted a microelectrode into the primary visual cortex of an anesthetized cat, and eavesdropped on a single neuron.
They could neither monitor nor control the activity of any other neurons in the network -- they could only listen to one neuron at a time.  What they {\it could} control was the visual stimulus.  In an attempt to get the neuron to fire, they projected black and white patterns on a screen in front of the open-eyed cat.  Remarkably, they found that the neuron they were listening to fired rapidly when the screen showed a black bar at a certain angle -- say, 45$^\circ$.  Other neurons responded to different angles.  It was as though each neuron was a sensor for a particular feature of the visual scene.  Its activity could be predicted without knowing anything about the network, but by simply looking {\it outside} the cat's brain -- at the stimulus on the screen.  

\begin{figure}[!h]
\includegraphics[width = 5in]{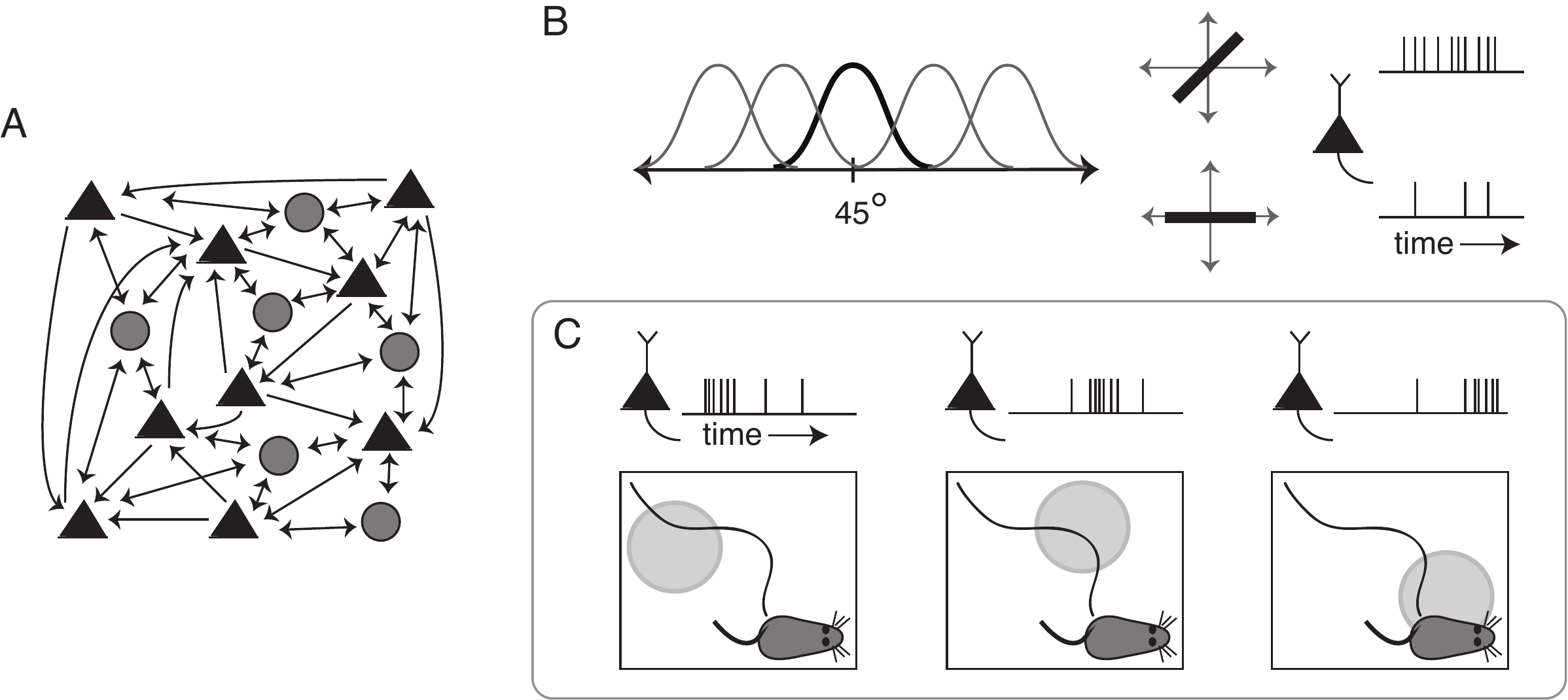}
\caption{\small The neural network and neural coding pictures.  (A) Pyramidal neurons (triangles) are embedded in a recurrent network together with inhibitory interneurons (circles).  (B) An orientation-tuned neuron in primary visual cortex with a preferred angle of 45$^\circ$.  The neuron fires many spikes in response to a bar at a 45$^\circ$ angle in the animal's visual field, but few spikes in response to a horizontal bar. (C) Place cells in the hippocampus fire when the animal passes through the corresponding place field.  The activity of three different neurons is shown (top), while the animal traces a trajectory starting at the top left corner of its environment (bottom).  Each neuron's activity is highest when the animal passes through the corresponding place field (shaded disc).}
\label{fig:net-RF}
\end{figure}

Hubel and Wiesel had discovered orientation-tuned neurons \cite{HubelWiesel59}, whose collective activity comprises a {\it neural code} for angles in the visual field (see Figure~\ref{fig:net-RF}B).  Although they inhabit a large, densely-connected cortical network, these neurons do not behave as unpredictable units governed by complicated dynamics.  Instead, they appear to be responding directly to stimuli in the outside world.  Their activity has {\it meaning}.

A decade later, O'Keefe made a similar discovery, this time involving neurons in a different area of the brain -- the hippocampus.  Unlike the visual cortex, there is no obvious sensory pathway to the hippocampus.  This made it all the more mysterious when O'Keefe reported that his neurons were responding selectively to different locations in the animal's physical environment \cite{OKeefe}.  These neurons, dubbed {\it place cells}, act as position sensors in space.  
When an animal is exploring a particular environment, a place cell increases its firing rate as the animal passes through its corresponding {\it place field} -- that is, the localized region to which the neuron preferentially responds (see Figure~\ref{fig:net-RF}C).

Like Hubel and Wiesel, who received a Nobel prize for their work in 1981 \cite{nobelprize81}, O'Keefe's discovery of place cells had an enormous impact in neuroscience.   In 2014, he shared the Nobel prize with Edvard and May-Britt Moser \cite{nobelprize14}, former postdocs of his who went on to discover an even stranger class of neurons that encode position, in a neighboring area of hippocampus called the entorhinal cortex.  These neurons, called {\it grid cells}, display periodic place fields that are arranged in a hexagonal lattice.  We'll come back to grid cells in the next section.

So, are neurons nodes in a network? or autonomous sensors of the outside world?
Both pictures are valid, and yet they lead to very different models of neural behavior.  Neural network theory deals with the first picture, and seeks to understand how the activity of neurons emerges from properties of the network.  In contrast, neural coding theory often treats the network as a black box, focusing instead on the relationship between neural activity
and external stimuli.  Many of the most interesting problems in neuroscience are about {\it understanding the neural code}.  This includes, but is not limited to, figuring out the basic principles by which neural activity represents sensory inputs to the eyes, nose, ears, whiskers, and tongue.  Because of the discoveries of Hubel and Wiesel, O'Keefe, and many others, we often know more about the coding properties of single neurons than we do about the networks to which they belong.   
But many open questions remain.  And topology, as it turns out, is a natural tool for understanding the neural code.

\section{Topology of hippocampal place cell codes}

\begin{figure}[!h]
\includegraphics[width = 5in]{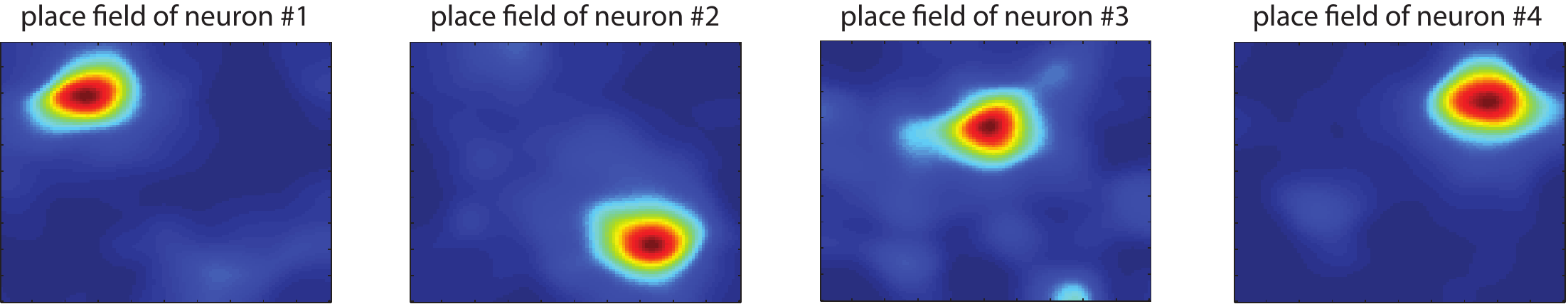}
\caption{\small Place fields for four place cells, recorded while a rat explored a 2-dimensional square box environment. Place fields were computed from data provided by the Pastalkova lab.}
\label{fig:place-fields}
\end{figure}

The term {\it hippocampal place cell code} refers to the neural code used by place cells in the hippocampus to encode the animal's position in space. 
Most of the research about place cells, including O'Keefe's original discovery, has been performed in rodents (typically rats), and the experiments typically involve an animal moving around in a restricted environment (see Figure~\ref{fig:net-RF}C).  It was immediately understood that a population of place cells, each having a different place field, could collectivity encode the animal's position in space \cite{OKeefe-book}, even though for a long time electrophysiologists could only monitor one neuron at a time.  When simultaneous recordings of place cells became possible, it was shown via statistical inference (using previously measured place fields) that the animal's position could indeed be inferred from population place cell activity \cite{brown1998statistical}.  Figure~\ref{fig:place-fields} shows four place fields corresponding to simultaneously recorded place cells in area CA1 of rat hippocampus.

The role of topology in place cell codes begins with a simple observation, which is perhaps obvious to anyone familiar with both place fields in neuroscience and elementary topology.  
First, let's recall the standard definitions of an open cover and a good cover.

\begin{definition}
Let $X$ be a topological space.  A collection of open sets, $\U = \{U_1,\ldots,U_n\}$, is an {\it open cover} of $X$ if $X = \bigcup_{i =1}^n U_i.$   We say that $\U$ is a {\it good cover} if every non-empty intersection $\bigcap_{i \in \sigma} U_i$, for $\sigma \subseteq \{1,\ldots, n\}$, is contractible.
\end{definition}

\begin{figure}[!h]
\includegraphics[width = 5in]{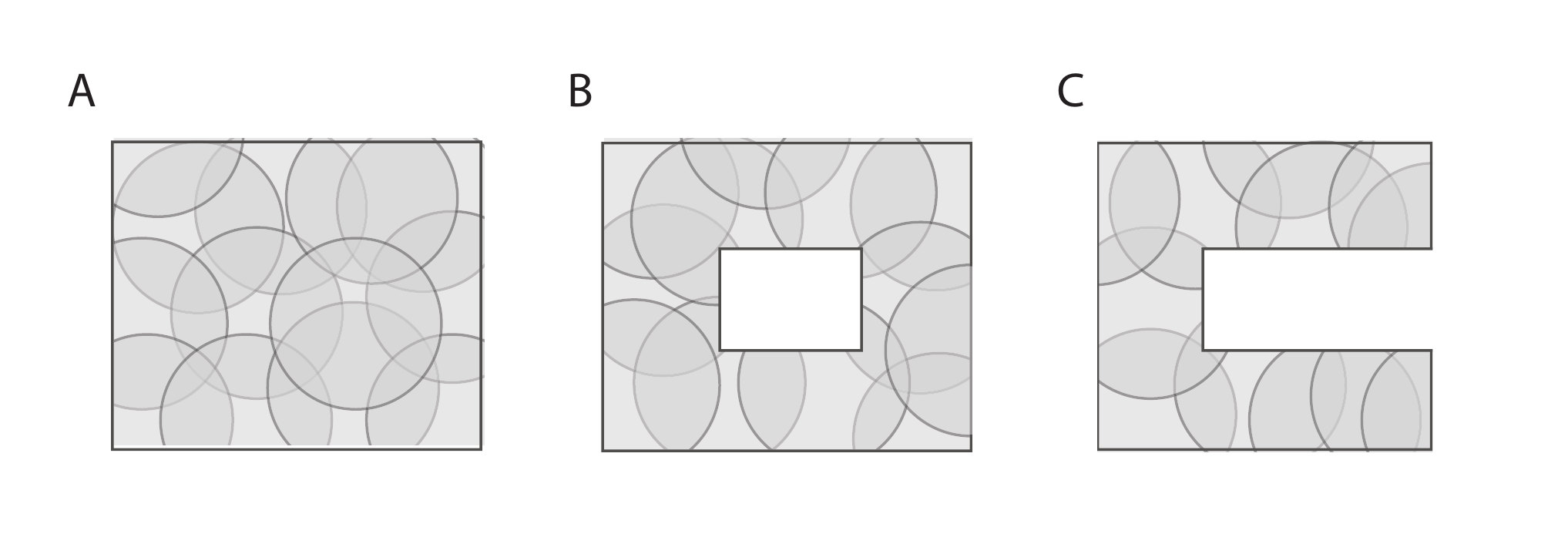}
\caption{\small Three environments for a rat: (A) A square box environment, also known as an ``open field''; (B) an environment with a hole or obstacle in the center;
and (C) a maze with two arms.  Each environment displays a collection of place fields (shaded discs) that fully cover the underlying space.}
\label{fig:environments}
\end{figure}

Next, observe that a collection of place fields in a particular environment looks strikingly like an open cover, with each $U_i$ corresponding a place field. 
Figure~\ref{fig:environments} displays three different environments, typical of what is used in hippocampal experiments with rodents, together with schematic arrangements of place fields in each.

Moreover, since place fields are approximately convex (see Figure~\ref{fig:place-fields}) it is not unreasonable to assume that they form a good cover of the underlying space.
This means the Nerve Lemma applies.  Recall the notion of the {\it nerve}\footnote{Note that the name ``nerve'' here predated any connection to neuroscience!} of a cover:
$$\N(\U) \od \{\sigma \subset [n] \mid \bigcap_{i \in \sigma} U_i \neq \emptyset \},$$
where $[n] = \{1,\ldots,n\}$.  Clearly, if $\sigma \in \N(\U)$ and $\tau \subset \sigma$, then $\tau \in \N(\U)$.  This property shows that $\N(\U)$ is an abstract {\it simplicial complex} on the vertex set $[n]$ -- that is, it is a set of subsets of $[n]$ that is closed under taking further subsets.  If $X$ is a sufficiently ``nice'' topological space, then the following well-known lemma holds.

\begin{lemma}[Nerve Lemma]\label{lemma:nerve}
Let $\U$ be a good cover of $X$.  Then $\N(\U)$ is homotopy-equivalent to $X$.  In particular, $\N(\U)$ and $X$ have exactly the same homology groups.
\end{lemma}

\begin{figure}[!h]
\includegraphics[width=4.5in]{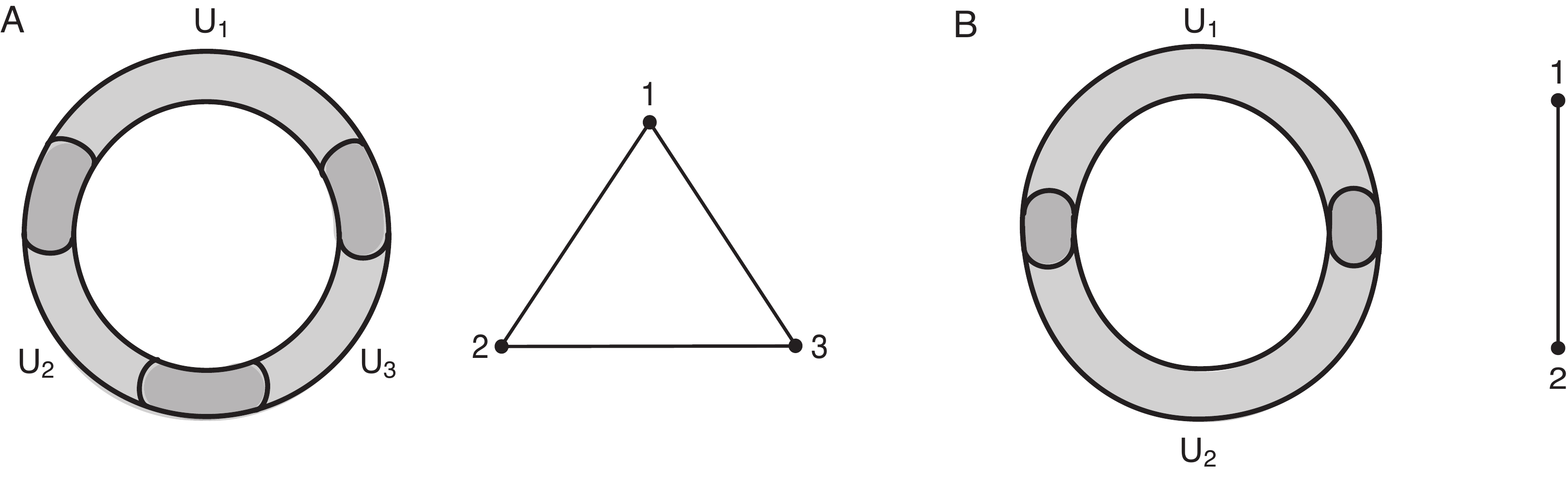}
\caption{Good and bad covers.  (A) A good cover $\U = \{U_1, U_2, U_3\}$ of an annulus (left), and the corresponding nerve $\N(\U)$ (right).  (B) A ``bad'' cover of the annulus (left), and the corresponding nerve (right).  Only the nerve of the good cover accurately reflects the topology of the annulus.}
\label{fig:good-bad-cover}
\end{figure}

It is important to note that the Nerve Lemma fails if the good cover assumption does not hold.  Figure~\ref{fig:good-bad-cover}A depicts a good cover of an annulus by three open sets.  The corresponding nerve (right) exhibits the topology of a circle, which is indeed homotopy-equivalent to the covered space.  In Figure~\ref{fig:good-bad-cover}B, however, the cover is {\it not} good, because the intersection $U_1 \cap U_2$ consists of two disconnected components, and is thus not contractible.  Here the nerve (right) is homotopy-equivalent to a point, in contradiction to the topology of the covered annulus.

\begin{figure}[!h]
\includegraphics[width = 1.5in]{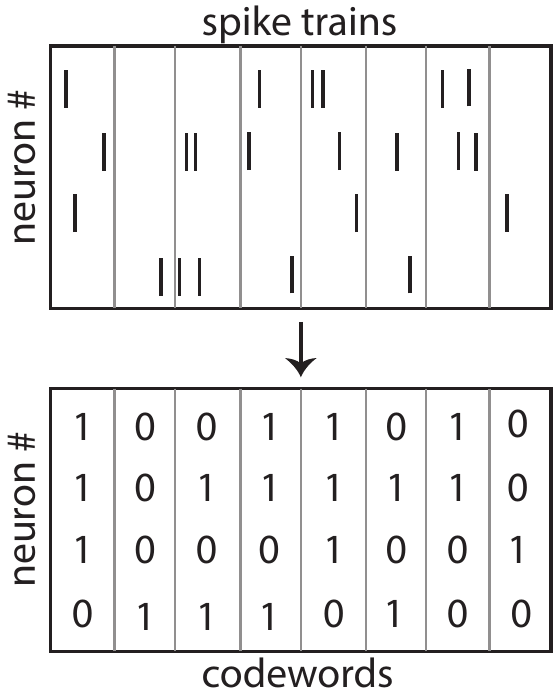}
\caption{\small By binning spike trains for a population of simultaneously-recorded neurons, one can infer subsets of neurons that co-fire.  If these neurons were place cells,
then the first codeword 1110 indicates that $U_1\cap U_2 \cap U_3 \neq \emptyset$, while the third codeword 0101 tells us $U_2 \cap U_4 \neq \emptyset$.}
\label{fig:spikes2codewords}
\end{figure}

The wonderful thing about the Nerve Lemma, when interpreted in the context of hippocampal place cells, is that $\N(\U)$ can be inferred from the activity of place cells alone -- without actually knowing the place fields $\{U_i\}$.  This is because the concurrent activity of a group of place cells, indexed by $\sigma \subset [n]$, indicates that the corresponding place fields have a non-empty intersection: $\bigcap_{i \in \sigma} U_i \neq \emptyset$.  
In other words, if we were eavesdropping on the activity of a population of place cells as the animal fully explored its environment, then by finding which subsets of neurons co-fire (see Figure~\ref{fig:spikes2codewords}) we could in principle estimate $\N(\U)$, even if the place fields themselves were unknown.  Lemma~\ref{lemma:nerve} tells us that the homology of the simplicial complex $\N(\U)$ precisely matches the homology of the environment $X$.  The place cell code thus naturally reflects the topology of the represented space.\footnote{In particular, place cell activity from the environment in Figure~\ref{fig:environments}B could be used to detect the non-trivial first homology group of the underlying space, and thus distinguish this environment from that of Figure~\ref{fig:environments}A or~\ref{fig:environments}C.}

These and related observations have led some researchers to speculate that the hippocampal place cell code is fundamentally topological in nature \cite{dabaghian2012topological, chen2014neural}, while others (including this author) have argued that considerable geometric information is also present and can be extracted using topological methods \cite{curto2008cell, giusti2015clique}.  In order to disambiguate topological and geometric features, Dabaghian et. al. performed an elegant experiment using linear tracks with flexible joints  \cite{dabaghian2014reconceiving}.  This allowed them to alter geometric features of the environment, while preserving the topological structure as reflected by the animal's place fields.  They found that place fields recorded from an animal running along the morphing track moved together with the track, preserving the relative sequence of locations despite changes in angles and movement direction.  In other words, the place fields respected topological aspects of the environment more than metric features \cite{dabaghian2014reconceiving}.

\begin{figure}[!h]
\includegraphics[width = 5in]{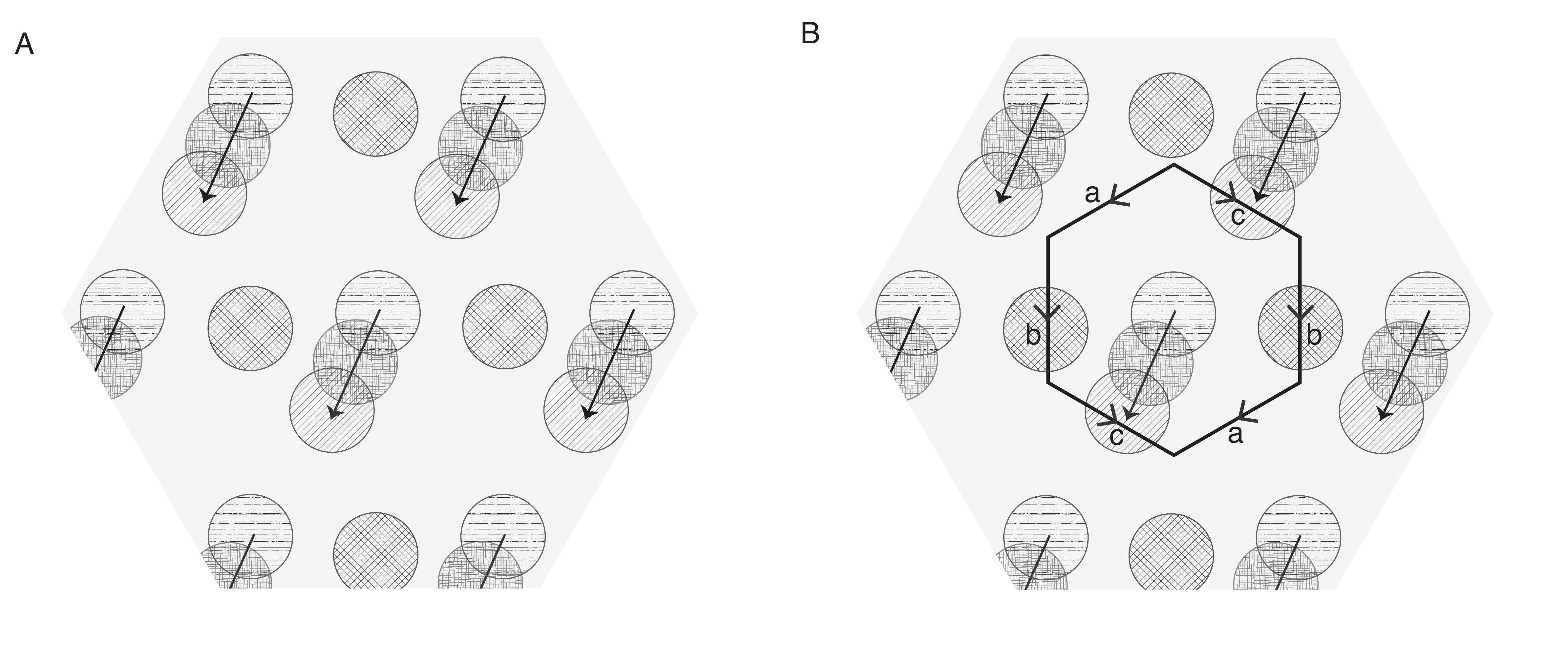}
\caption{\small Firing fields for grid cells. (A) Firing fields for four entorhinal grid cells.  Each grid field forms a hexagonal grid in the animal's two-dimensional environment, and each field thus has multiple disconnected regions.  (B) A hexagonal fundamental domain contains just one disc-like region per grid cell.  Pairs of edges with the same label (a, b, or c) are identified, with orientations specified by the arrows.}
\label{fig:grid-cells}
\end{figure}

What about the entorhinal grid cells?  These neurons have firing fields with multiple disconnected components, forming a hexagonal grid (see Figure~\ref{fig:grid-cells}A).  This means that grid fields violate the good cover assumption of the Nerve Lemma -- if we consider them as an open cover for the entire 2-dimensional environment.  If, instead, we restrict 
attention to a fundamental domain for these firing fields, as illustrated in Figure~\ref{fig:grid-cells}B, then each grid field has just one (convex) component, and the Nerve Lemma applies.  From the spiking activity of grid cells we could thus infer the topology of this fundamental domain.  The reader familiar with the topological classification of surfaces may recognize that this hexagonal domain, with the identification of opposite edges, is precisely a torus.  To see this, first identify the edges labeled ``a'' to get a cylinder.  Next, observe that the boundary circles on each end of the cylinder consist of the edges ``b'' and ``c'', but with a 180$^\circ$ twist between the two ends.  By twisting the cylinder, the two ends can be made to match so that the ``b'' and ``c'' edges get identified.  This indicates that the space represented by grid cells is not the full environment, but a torus.

\section{Topology in neuroscience: a bird's-eye view}
The examples from the previous section are by no means the only way that topology is being used in neuroscience.  Before plunging into further details about what topology can tell us about neural codes, we now pause for a moment to acknowledge some other interesting applications.  The main thing they all have in common is their recency.  This is no doubt due to the rise of computational and applied algebraic topology, a relatively new development in applied math that was highlighted in the Current Events Bulletin nearly a decade ago 
\cite{ghrist2007barcodes}.

Roughly speaking, the uses of topology in neuroscience can be categorized into three (overlapping) themes: (i) ``traditional'' topological data analysis applied to neuroscience; (ii) an upgrade to network science; and (iii) understanding the neural code.  Here we briefly summarize work belonging to (i) and (ii).  In the next section we'll return to (iii), which is the main focus of this talk.

\subsection{``Traditional'' TDA applied to neuroscience data sets.}  The earliest and most familiar applications of topological data analysis (TDA) focused on the problem of estimating the ``shape'' of point-cloud data.  This kind of data set is simply a collection of points, $x_1,\ldots,x_\ell \in \RR^n$, where $n$ is the dimensionality of the data.  A question one could ask is: do these points appear to have been sampled from a lower-dimensional manifold, such as a torus or a sphere?  The strategy is to consider open balls $B_\varepsilon(x_i)$ of radius $\varepsilon$ around each data point, and then to construct a simplicial complex $\mathcal{K}_\varepsilon$ that captures information about how the balls intersect.  This simplicial complex can either be the Cech complex (i.e., the nerve of the open cover defined by the balls), or the Vietoris-Rips complex (i.e., the clique complex of the graph obtained from pairwise intersections of the balls).  By varying $\varepsilon$, one obtains a sequence of nested simplicial complexes $\{\mathcal{K}_\varepsilon\}$ together with natural inclusion maps.  Persistent homology tracks homology cycles across these simplicial complexes, and allows one to determine whether there were homology classes that ``persisted'' for a long time.  For example, if the data points were sampled from a 3-sphere, one would see a persistent 3-cycle. 

There are many excellent reviews of persistent homology, including \cite{ghrist2007barcodes}, so I will not go into further details here.  Instead, it is interesting to note that one of the early applications of these techniques was in neuroscience, to analyze population activity in primary visual cortex \cite{singh2008topological}.   Here it was found  that the topological structure of activity patterns is similar between spontaneous and evoked activity, and consistent with the topology of a two-sphere.  Moreover, the results of this analysis were interpreted in the context of neural coding, making this work exemplary of both themes (i) and (iii).  Another application of persistent homology to point cloud data in neuroscience was the analysis of the spatial structure of afferent neuron terminals in crickets \cite{brown2012structure}.  Again, the results were interpreted in terms of the coding properties of the corresponding neurons, which are sensitive to air motion detected by thin mechanosensory hairs on the cricket.  Finally, it is worth mentioning that these types of analyses are not confined to neural activity.  For example, in  \cite{bendich2014persistent} the statistics of persistent cycles were used to study brain artery trees.

\subsection {An upgrade to network science.}  There are many ways of constructing networks in neuroscience, but the basic model that has been used for all of them is the graph.  The vertices of a graph can represent neurons, cell types, brain regions, or fMRI voxels, while the edges reflect interactions between these units.  Often, the graph is weighted and the edge weights correspond to correlations between adjacent nodes.  For example, one can model a functional brain network from fMRI data as a weighted graph where the edge weights correspond to activity correlations between pairs of voxels.  At the other extreme, a network where the vertices correspond to neurons could have edge weights that reflect either pairwise correlations in neural activity, or synaptic connections.

Network science is a relatively young discipline that focuses on analyzing networks, primarily using tools derived from graph theory.  The results of a particular analysis could range from determining the structure of a network to identifying important subgraphs and/or graph-theoretic statistics (the distribution of in-degree or out-degree across nodes, number of cycles, etc.) that carry meaning for the network at hand.  Sometimes, graph-theoretic features do not carry obvious meaning, but are nevertheless useful for distinguishing networks that belong to distinct classes.  For example, a feature could be characteristic of functional brain networks derived from a subgroup of subjects, distinguishing them from a ``control'' group.
In this way graph features may be a useful diagnostic tool for distinguishing diseased states, pharmacologically-induced states, cognitive abilities, or uncovering systematic differences based on gender or age.

\begin{figure}[!h]
\includegraphics[width=4.5in]{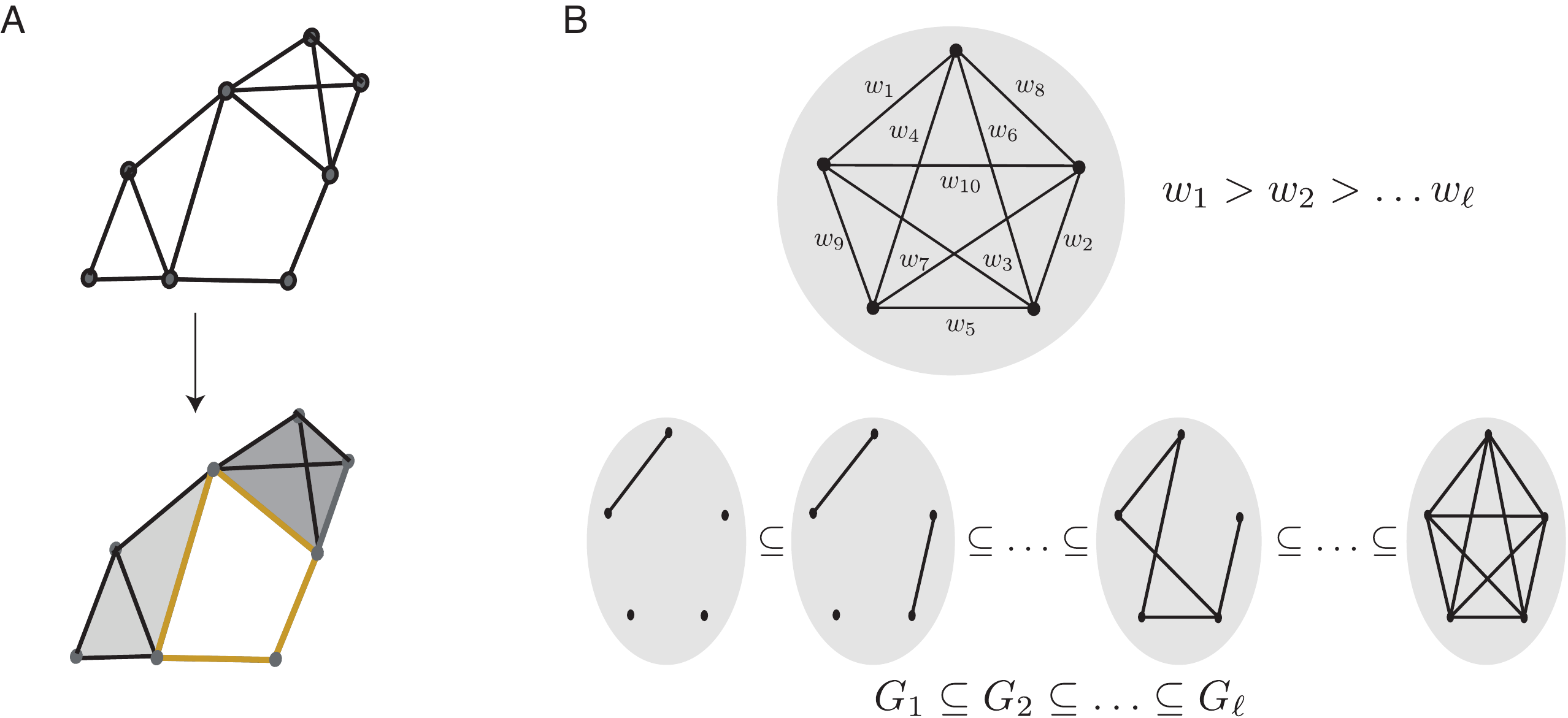}
\caption{Network science models: from graphs to clique complexes and filtrations.}
\label{fig:graph-cliques}
\end{figure}

The recent emergence of topological methods in network science stems from the following ``upgrade'' to the network model: instead of a graph, one considers a simplicial complex.  Sometimes this simplicial complex reflects higher-order interactions that are obtained from the data, and sometimes it is just the {\it clique complex} of the graph $G$:
$$X(G) = \{\sigma \subset [n] \mid (ij) \in G \text{ for all } i,j \in \sigma\}.$$
In other words, the higher-order simplices correspond to cliques (all-to-all connected subgraphs) of $G$.
Figure~\ref{fig:graph-cliques}A shows a graph (top) and the corresponding clique complex (bottom), with shaded simplices  corresponding to two 3-cliques and a 4-clique.  The clique complex fills in many of the 1-cycles in the original graph, but some 1-cycles remain (see the gold 4-gon), and higher-dimensional cycles may emerge.  Computing homology groups for the clique complex is then a natural way to detect topological features that are determined by the graph.  In the case of a weighted graph, one can obtain a sequence of clique complexes by considering a related sequence of simple graphs, where  each graph is obtained from the previous one by adding the edge corresponding to the next-highest weight (see Figure~\ref{fig:graph-cliques}B).  The corresponding sequence of clique complexes, $\{X(G_i)\}$, can then be analyzed using persistent homology.  Other methods for obtaining a sequence of simplicial complexes from a network are also possible, and may reflect additional aspects of the data such as the temporal evolution of the network.

For a more thorough survey of topological methods in network science, I recommend the forthcoming review article \cite{giusti2015two}.  Here I will only mention
that topological network analyses have already been used in a variety of neuroscience applications, many of them medically-motivated: fMRI networks in patients with ADHD \cite{ellis2014describing}; FDG-PET based networks in children with autism and ADHD \cite{lee2011discriminative};  morphological networks in deaf adults \cite{kim2014morphological}; metabolic connectivity in epileptic rats \cite{choi2014abnormal}; and functional EEG connections in depressed mice \cite{khalid2014tracing}.
Other applications to fMRI data include human brain networks during learning \cite{stolz2014computational} and drug-induced states \cite{petri2014homological}.  At a finer scale, recordings of neural activity can also give rise to functional connectivity networks among neurons (which are not the same as the neural networks defined by synaptic connections).  These networks have also been analyzed with topological methods \cite{pirino2014topological, giusti2015clique, spreemann2015using}.

\section{The code of an open cover}
We now return to neural codes.  We have already seen how the hippocampal place cell code reflects the topology of the underlying space, via the nerve $\N(\U)$ of a place field cover.  In this section, we will associate a binary code to an open cover.  This notion is closer in spirit to a combinatorial neural code (see Figure~\ref{fig:spikes2codewords}), and carries more detailed information than the nerve.  In the next section, we'll see how topology is being used to determine intrinsic features of neural codes, such as convexity and dimension.

First, a few definitions.   A {\it binary pattern} on $n$ neurons is a string of $0$s and $1s$, with a $1$ for each active neuron and a $0$ denoting silence; 
equivalently, it is a subset of (active) neurons 
$\sigma \subset [n].$  (Recall that $[n] = \{1,\ldots,n\}$.)
We use both notations interchangeably.  For example, $10110$ and $\sigma = \{1,3,4\}$ refer to the same pattern, or codeword, on $n = 5$ neurons.  A {\it combinatorial neural code} on $n$ neurons is a collection of binary patterns $\C \subset 2^{[n]}$.  In other words, it is a binary code of length $n$, where we interpret each binary digit as the ``on'' or ``off'' state of a neuron.
The {\it simplicial complex of a code}, $\Delta(\C)$, is the smallest abstract simplicial complex on $[n]$ that contains all elements of $\C$.  In keeping with the hippocampal place cell example, we are interested in codes that correspond to open covers of some topological space.

\begin{figure}[!h]
\includegraphics[width=5in]{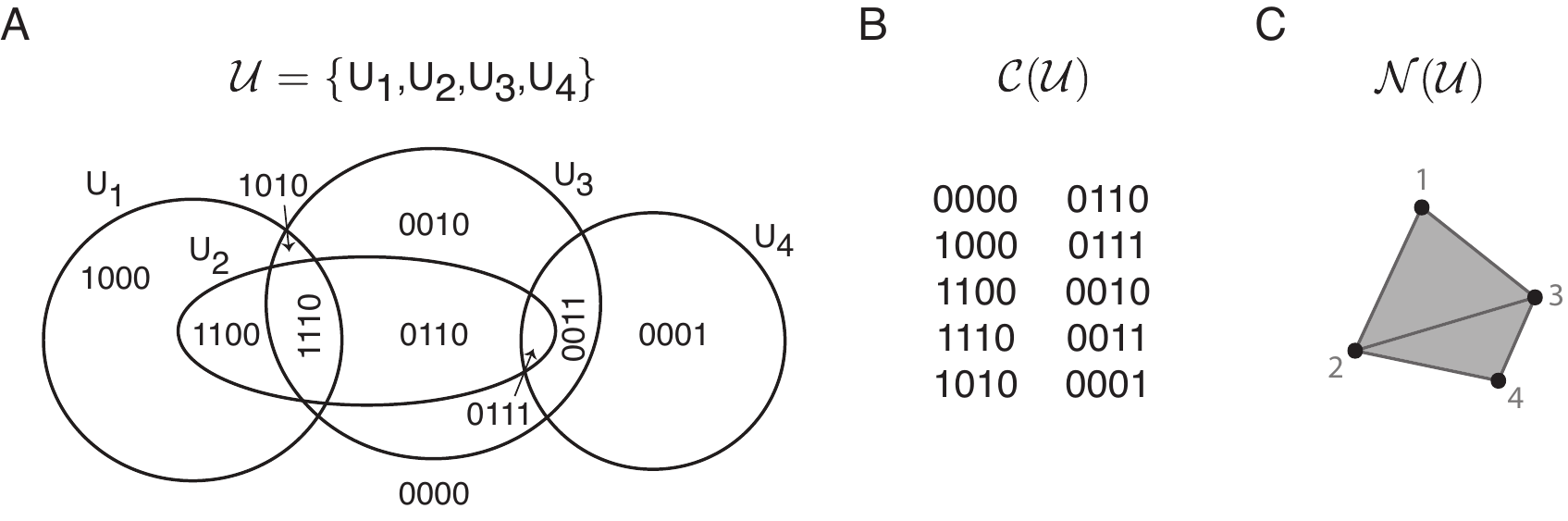}
\caption{Codes and nerves of open covers.  (A) An open cover $\U$, with each region carved out by the cover labeled by its corresponding codeword.  (B) The code $\C(\U)$.  (C) The nerve $\N(\U)$.}
\label{fig:code-cover}
\end{figure}

\begin{definition}
Given an open cover $\U$, the {\it code of the cover} is the combinatorial neural code
$$\C(\U) \od \{ \sigma \subset [n] \mid \bigcap_{i \in \sigma} U_i \setminus \bigcup_{j \in [n] \setminus \sigma} U_j \neq \emptyset \}.$$ 
\end{definition}

Each codeword in $\C(\U)$ corresponds to a region that is defined by the intersections of the open sets in $\U$ (Figure~\ref{fig:code-cover}A).
Note that the code $\C(\U)$ is not the same as the nerve $\N(\U)$.  Figures~\ref{fig:code-cover}B and ~\ref{fig:code-cover}C display the code and the nerve of the open cover in Figure~\ref{fig:code-cover}A.  While the nerve encodes which subsets of the $U_i$s have non-empty intersections, the code
also carries information about set containments.  For example, the fact that $U_2 \subseteq U_1 \cup U_3$ can be inferred from $\C(\U)$ because each codeword of the form
$*1**$ has an additional $1$ in position 1 or 3, indicating that if neuron 2 is firing then so is neuron 1 or 3.  Similarly, the fact that $U_2 \cap U_4 \subseteq U_3$ can be 
inferred from the code because any word of the form $*1*1$ necessarily has a 1 in position 3 as well.  These containment relationships go beyond simple intersection data, and cannot be obtained from the nerve $\N(\U)$.  On the other hand, the nerve can easily be recovered from the code since $\N(\U)$ is the smallest simplicial complex that contains it -- that is, 
$$\N(\U) = \Delta(\C(\U)).$$  
$\C(\U)$ thus carries more detailed information than what is available in $\N(\U)$.  The combinatorial data in $\C(\U)$ can also be encoded algebraically via the {\it neural ideal} \cite{neural_ring}, much as simplicial complexes are algebraically encoded by Stanley-Reisner ideals \cite{miller-sturmfels}.

It is easy to see that any binary code, $\C \subseteq \{0,1\}^n$, can be realized as the code of an open cover.\footnote{For example,
if the size of the code is $|\C| = \ell$, we could choose disjoint open intervals $B_1, \ldots, B_\ell \subset \RR$, one for each codeword, and define
the open sets $U_1,\ldots,U_n$ such that $U_i$ is the union of all open intervals $B_j$ corresponding to codewords in which neuron $i$ is ``on'' (that is, there is a $1$ in position $i$ of the codeword).  Such a cover, however, consists of highly disconnected sets and its properties reflect very little of the underlying space -- in particular, the good cover assumption of the Nerve Lemma is violated.}  It is not true, however, that any code can arise from a good cover or a {\it convex cover}  -- that is, an open cover consisting of convex sets.  The following lemma illustrates the simplest example of what can go wrong.

\begin{lemma}\label{lemma:disconnect}
Let $\C \subset \{0,1\}^3$ be a code that contains the codewords $110$ and $101$, but does \underline{not} contain $100$ and $111$.  Then $\C$ is not the code of a good or convex cover.
\end{lemma}

The proof is very simple.  Suppose  $\U = \{U_1, U_2, U_3\}$ is a cover such that $\C = \C(\U)$. Because neuron $2$ or $3$ is ``on'' in any codeword for which neuron $1$ is ``on,''
we must have that $U_1 \subset U_2 \cup U_3$.  Moreover, we see from the code that $U_1 \cap U_2 \neq \emptyset$ and $U_1 \cap U_3 \neq \emptyset$, while $U_1 \cap U_2 \cap U_3 = \emptyset$.  This means we can write $U_1$ as a disjoint union of two non-empty sets: $U_1 = (U_1 \cap U_2) \cup (U_1 \cap U_3).$  $U_1$ is thus disconnected, and hence $\U$ can be neither a good nor convex cover. 

\section{Using topology to study intrinsic properties\\ of neural codes}
 
In our previous examples from neuroscience, the place cell and grid cell codes can be thought of as arising from convex sets covering an underlying space.  
Because the spatial correlates of these neurons are already known, it is not difficult to infer what space is being represented by these codes.  What could we say
if we were given just a code, $\C \subset \{0,1\}^n$, without {\it a priori} knowledge of what the neurons were encoding?  Could we tell whether such a code
can be realized via a convex cover? 

\subsection{What can go wrong.}

As seen in Lemma~\ref{lemma:disconnect}, not all codes can arise from convex covers.  Moreover, the problem that prevents the code in Lemma~\ref{lemma:disconnect} from being convex is topological in nature.  Specifically, what happens in the example of Lemma~\ref{lemma:disconnect} is that the code dictates there must be a set containment, 
$$U_\sigma \subseteq \bigcup_{j \in \tau} U_j,$$
where $U_\sigma = \bigcap_{i \in \sigma} U_i$, but the nerve of the resulting cover of $U_\sigma$ by the sets $\{U_\sigma \cap U_j\}_{j \in \tau}$ is not contractible.
This leads to a contradiction if the sets $U_i$ are all assumed to be convex, because the sets $\{U_\sigma \cap U_j\}_{j \in \tau}$ are then also convex and thus form a good cover of $U_\sigma$.  Since $U_\sigma$ itself is convex, and the Nerve Lemma holds, it follows that $\N(\{U_\sigma \cap U_j\}_{j \in \tau})$ must be contractible, contradicting the data of the code.

These observations lead to the notion of a {\it local obstruction} to convexity \cite{no-go}, which captures the topological problem that arises if certain codes
are assumed to have convex covers.  The proof of the following lemma is essentially the argument outlined above.

\begin{lemma}[\cite{no-go}] \label{lemma:no-obs}
If $\C$ can be realized by a convex cover, then $\C$ has no local obstructions.
\end{lemma}

The idea of using local obstructions to determine whether or not a neural code has a convex realization has been recently followed up in a series of papers \cite{MRC, counterexample, intersection-complete}.  In particular, local obstructions have been characterized in terms of links, $\mathrm{Lk}_{\Delta}(\sigma)$, corresponding to ``missing'' codewords that are not in the code, but are elements of the simplicial complex of the code.

\begin{theorem}[\cite{MRC}] \label{thm:loc-obs}
Let $\C$ be a neural code, and let $\Delta = \Delta(\C)$.  Then $\C$ has no local obstructions if and only if $\mathrm{Lk}_{\Delta}(\sigma)$ is contractible for all $\sigma \in \Delta \setminus \C$.
\end{theorem}

It was believed, until very recently, that the converse of Lemma~\ref{lemma:no-obs} might also be true.  However, in \cite{counterexample} the following counterexample was discovered, showing that this is not the case.  Here the term {\it convex code} refers to a code that can arise from a convex open cover.

\begin{example}[\cite{counterexample}]
The code $\C = \{2345, 123, 134, 145, 13, 14, 23, 34, 45, 3, 4\}$ is not a convex code, despite the fact that it has no local obstructions.  
\end{example}

That this code has no local obstructions can be easily seen using Theorem~\ref{thm:loc-obs}.  The fact that there is no convex open cover, however, relies on convexity arguments that are not obviously topological.  Moreover, this code does have a good cover \cite{counterexample}, suggesting the existence of a new class of obstructions to convexity which may or may not be topological in nature.

\subsection{What can go right.}
Finally, it has been shown that several classes of neural codes are guaranteed to have convex realizations.  {\it Intersection-complete codes} satisfy the property that for any $\sigma, \tau \in \C$ we also have $\sigma \cap \tau \in \C$.  These codes (and some generalizations) were shown constructively to have convex covers in \cite{intersection-complete}.  Additional classes of codes with convex realizations have been described in \cite{MRC}.  

Despite these developments, a complete characterization of convex codes is still lacking.  Finding the minimum dimension needed for a convex realization is also an open question.

\section{Codes from networks}  

We end by coming back to the beginning.
Even if neural codes give us the illusion that neurons in cortical and hippocampal areas are directly sensing the outside world, we know that of course they are not.  Their activity patterns are shaped by the networks in which they reside.  What can we learn about the architecture of a network by studying its neural code?  This question requires an improved understanding of neural networks, not just neural codes.  While many candidate architectures have been proposed to explain, say, orientation-tuning in visual cortex, the interplay of neural network theory and neural coding is still in early stages of development.

\begin{figure}[!h]
\includegraphics[width=3in]{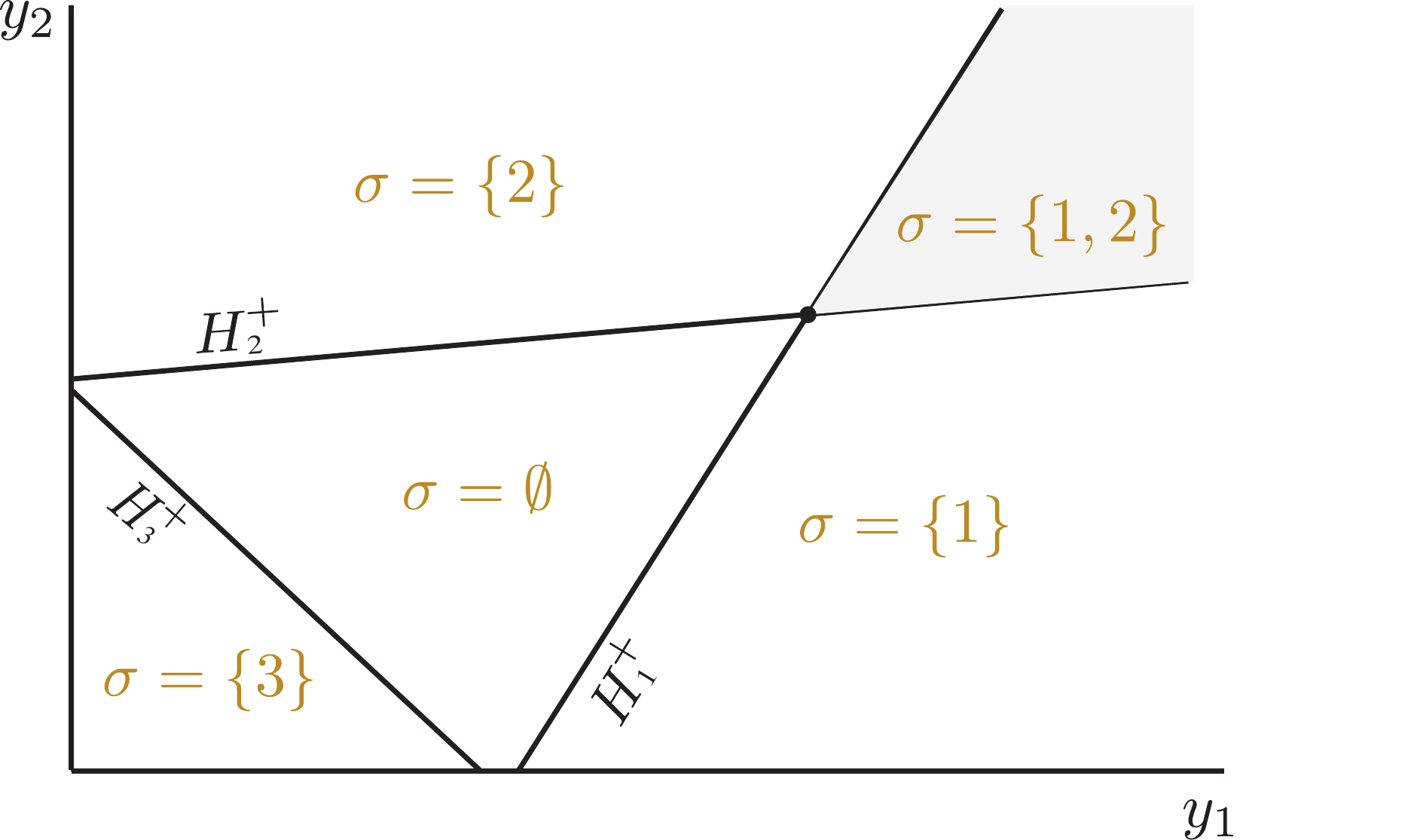}
\caption{A hyperplane arrangement in the positive orthant, and the corresponding feedforward code.}
\label{fig:FF-code}
\end{figure}

Perhaps the simplest example of how the structure of a network can constrain the neural code is the case of simple feedforward networks.  These networks have a single input layer of neurons, and a single output layer.  The resulting codes are derived from hyperplane arrangements in the positive orthant of $\RR^k$, where $k$ is the number of neurons in the input layer and each hyperplane corresponds to a neuron in the output layer (see Figure~\ref{fig:FF-code}).  Every codeword in a {\it feedforward code} corresponds to a chamber in such a hyperplane arrangement.  

It is not difficult to see from this picture that all feedforward codes are realizable by convex covers -- specifically, they arise from overlapping half-spaces \cite{no-go}.  On the other hand, not every convex code is the code of a feedforward network \cite{vladimir2015personal}.  Moreover, the discrepancy between feedforward codes and convex codes is not due to restrictions on their simplicial complexes.  As was shown in \cite{no-go}, every simplicial complex can arise as $\Delta(\C)$ for a feedforward code.  As with convex codes, a complete characterization of feedforward codes is still unknown.  It seems clear, however, that topological tools will play an essential role.

\section{Acknowledgments}

I would like to thank Chad Giusti for his help in compiling a list of references for topology in neuroscience.
I am especially grateful to Katie Morrison for her generous help with the figures.

\bibliographystyle{amsplain}
\bibliography{current-events}

\end{document}